
\magnification=\magstep1
\hfuzz=4truept
\nopagenumbers
\font\tif=cmr10 scaled \magstep3

\rightline{PUPT-1549$|$hep-lat/9507007}
\vfil
\centerline{\tif Improving lattice perturbation theory}
\vfil
\centerline{{\rm Vipul
Periwal}\footnote{${}^\dagger$}{vipul@puhep1.princeton.edu}}
\bigskip
\centerline{Department of Physics}\centerline{Princeton University
}\centerline{Princeton, New Jersey 08544-0708}
\vfil
\par\noindent Lepage and Mackenzie have shown that tadpole renormalization
and systematic improvement of lattice perturbation theory can lead to
much improved numerical results in lattice gauge theory.
It is shown that lattice perturbation theory using the Cayley parametrization
of unitary matrices gives a simple analytical approach to tadpole
renormalization, and that the Cayley parametrization gives lattice gauge
potentials gauge transformations close to the continuum form.  For example,
at the lowest order in perturbation theory, for SU(3) lattice
gauge theory, at $\beta=6,$ the `tadpole renormalized' coupling
$\tilde g^2 = {4\over 3} g^2,$ to be compared to the
non-perturbative numerical  value $\tilde g^2 = 1.7 g^2.$
\medskip
\vfil\eject

\def\part{\partial}
\def\wil{1} 
\def\wz{2}
\def\gross{3}
\def\z{4}
\def\spglcs{5}
\def\has{6}
\def\eps{\varepsilon}
\def\tr{\hbox{tr}}
\def\Lam{\Lambda}
\def\hcE{{\widehat\cE}}
\centerline{1. Introduction}
\bigskip
Lepage and Mackenzie[\wil]
 have made considerable advances in numerical simulations
of lattice gauge theories by improving the lattice gauge theory action
in two ways:
\item{1.} The action is improved in the sense of Symanzik's[\wz] program.
\item{2.} The action is improved by `tadpole renormalization'---a procedure
by which the gauge potential associated to each link variable
is given not by the usual correspondence
$$ U_\mu \leftrightarrow \exp(igaA_\mu)$$
but by
$$ U_\mu \leftrightarrow u_0 (1+igaA_\mu).$$
\par\noindent It is observed by Lepage and Mackenzie[\wil]
(and collaborators[\z])
that these two improvements together lead to a much better convergence
to a continuum limit, with orders of magnitude less computational effort.

Given these remarkable results, it is important to understand exactly what
is analytically involved in these improvements, especially for
tadpole renormalization, which is not as well-understood as the
Symanzik-type improvements.  One might hope that a better understanding of
the continuum limit would be useful as well for understanding how to link
the strong coupling expansion and continuum perturbation theory[\spglcs],
for example.

That the Lepage-Mackenzie prescription needs some explication can be seen,
for instance, by considering the gauge transformation properties of the
above correspondence.  A prime requirement for
gauge invariance is $UU^\dagger=1,$ which is not consistent with
$U_\mu \leftrightarrow u_0 (1+igaA_\mu)$ unless $u_0^{-2}= 1+(ga)^2A_\mu^2.$
The intuition is that $u_0$ incorporates short distance
fluctuations of the link variable, leaving $A_\mu$ as the continuum
gauge field at long distances.  However, it is obvious that the
exponential parametrization that is manifestly unitary cannot be
factorized in any simple manner for non-Abelian gauge theories, so as
to lead to the required form.  One is lead, then, to ask: Given that
tadpole renormalization does appear to work, what is the systematic
analytical approximation that underlies it?  Clearly, from the nomenclature,
one hopes to resum some subsets of graphs in lattice perturbation theory,
so as to make the rest of the lattice perturbation theory closer to
continuum perturbation theory. Consistent resummations of selected subsets of
graphs in continuum relativistic gauge theories are rare.  On the
lattice, of course, one has a great deal more freedom, and reorganizing
the perturbation theory to make the approach to the continuum simpler is
a good thing to consider.  The basic idea, then,  is that the reorganized
lattice
perturbation theory should be closer to continuum perturbation theory.

I show, in the present work, that the classical Cayley parametrization of
group elements has two things to recommend it over the exponential
parametrization in common use:
\item{I.} This parametrization leads to gauge transformations for
the lattice gauge potentials associated with links, that are `closer' to
continuum gauge transformations, than are the gauge transformations for
the potentials obtained via the exponential map.
\item{II.} In this parametrization, it is natural to reorganize the
lattice perturbation theory in a manner that eliminates certain tadpoles,
and renormalizes the na\"\i ve lattice coupling constant in the
way posited by Lepage and Mackenzie.  I calculate the background field
effective action in this parametrization after a suitable resummation, and
show that some care is needed in interpreting the resulting expansion
parameter.
\bigskip
\centerline{2. Gauge transformations}
\bigskip
The Cayley parametrization of a unitary matrix is
$$U \equiv{1+igA/2\over 1-igA/2}, \qquad A\ \ {\rm Hermitian}.$$
This parametrization is valid so long as $\det(U+1)\not=0,$ so
we may use it in lattice perturbation theory, when we expect $U\approx1.$
So at the most simplistic level, one could arrive at a relation
close to the Lepage-Mackenzie form by expanding
$$U_\mu = {1+igaA_\mu/2\over 1-igaA_\mu/2} = {1\over1-c^2}\left[1+\Delta +
\Delta^2 + \dots\right] \left(1+igaA_\mu+\Delta (1-c^2)\right),$$
with $\Delta\equiv(c^2-g^2a^2A_\mu^2/4)/(c^2-1).$
This expansion is now consistent with unitary link variables.
The
series expansion will converge the best when inserted in correlation
functions if $c^2=-\langle g^2a^2A_\mu^2\rangle.$
The higher terms in the series will then involve insertions of
powers of $(c^2-g^2a^2A_\mu^2/4),$ and self-contractions (tadpoles) will
be cancelled.  The higher terms are, of course, crucial for unitary
link variables.  This particular form of resumming tadpoles is
likely not a good idea, since it is manifestly not gauge invariant.
The same idea can, however, be implemented at the level of the lattice gauge
theory action, as will be shown later in this paper.

Let us turn now to the question of  gauge invariance of the gauge
potential $A$ defined by the Cayley parametrization.  It will be useful
to contrast the lattice gauge invariance of this parametrization with that
of the usual exponential parametrization, so I first recall the standard
result.  (I have rescaled $A$ to absorb $g$ for this part of the discussion.)
If $U_\mu(x)\equiv \exp(iaA_\mu(x)),$ then gauge transformations map
$U_\mu$ to $\tilde U_\mu(x) \equiv \eta(x) U_\mu(x)\eta^\dagger(x+e_\mu).$
Let $\eta(x) = 1+i\eps(x),$ then
one has
$$\delta U_\mu(x) U^\dagger_\mu(x) = -iaD_\mu\eps(x),$$
with the lattice covariant derivative defined by
$$D_\mu\eps(x) \equiv {1\over a} \left(U_\mu(x) \eps(x+e_\mu)U^\dagger_\mu(x)
-\eps(x)\right).$$
We are interested in the change in $A_\mu,$ induced by such a change
in $U_\mu,$ so we expand both sides in powers of $A,$ to find
$$\eqalign{i\delta A_\mu(x) &-{a\over 2} \{A_\mu,\delta A_\mu(x)\}
-{ia^2\over 6}\big(\{\delta A_\mu(x),A_\mu(x)^2\}
+A_\mu(x)\delta A_\mu(x)A_\mu(x) \big) +\dots \cr
= &-i\left[\tilde D_\mu\eps(x) + {ia\over 2}\{A_\mu(x),\Delta_\mu\eps(x)\}
B
- {a\over 2}[A_\mu(x)^2,\eps(x)]+\dots\right], \cr}$$
where $\Delta_\mu\eps(x)\equiv \left(\eps(x+e_\mu)-\eps(x)\right)/a,$
and
$$\tilde D_\mu\eps(x)\equiv {1\over a}\left(\eps(x+e_\mu)-\eps(x)\right)
+{i\over 2}[A_\mu(x),\eps(x+e_\mu)+\eps(x)].$$
It follows then that the gauge transformations of the gauge potential
on the lattice,
defined by the exponential parametrization,
differ from the continuum form of the gauge transformations
at $O(a).$  Explicitly, if $\delta A_\mu = - \tilde D_\mu\eps(x) +a \varpi,$
then
$$\varpi = -{1\over 4} \{A_\mu,[A_\mu,\eps(x+e_\mu)+\eps(x)]\} +{1\over2}
[A_\mu(x)^2,\eps(x)] + O(a).$$

The same calculation for the Cayley parametrization is much simpler, and leads
in  a straightforward manner to
$$\delta A_\mu(x) = -\tilde D_\mu\eps(x)
- {a^2\over 4}A_\mu(x)\Delta_\mu \eps(x)A_\mu(x),$$
an exact result as opposed to the infinite series one obtains in the
exponential
parametrization.  Thus the Cayley parametrization is closer to the
continuum in the sense that gauge transformations of the Cayley gauge
potential differ {\it only} at $O(a^2),$ from $-\tilde D_\mu\eps(x) ,$
whereas gauge transformation for the exponential parametrization differ
at $O(a),$ with contributions at all higher powers of $a.$
Of course, $-\tilde D_\mu\eps(x)$ itself differs from the continuum
form at $O(a).$

One may wonder if it is possible to improve the lattice definition of
gauge potential even further, to make the deviation from the continuum
even higher order in $a.$  It is a simple matter to show that
parametrizations of the form $U_\mu = 1+f(iaA)/1+f(-iaA),$ with $f$ a
real polynomial, do not accomplish this.

\bigskip
\centerline{3. Tadpole resummation in the action}
\bigskip
Consider now using the Cayley parametrization in the Wilson action,
$$S_W = {1\over g^2} \sum_{\rm plaquettes} \Re \tr \left[1-
U_\mu(x)U_\nu(x+e_\mu) U_\mu(x+e_\nu)U_\nu(x)\right]
\equiv {1\over g^2} \sum_{\rm plaquettes} \Re
\tr \left[1-W_{\mu\nu}\right].$$
A particularly interesting question to address is the following: The
renormalization group scale parameter  for weak coupling lattice
perturbation
theory, $\Lambda_L,$ is very small[\has,\gross]
 compared to the continuum value, computed
with {\it e.g.} Pauli-Villars,
$${\Lam_L\over \Lam_{PV}} \approx 0.02, \qquad {\rm for\ \ SU}(\infty).$$
The fact that this ratio is small due to tadpole diagrams was
noted in the original calculation of Dashen and Gross[\gross]---this is one
of the motivations for tadpole renormalization according to Lepage and
Mackenzie.  Therefore, a good test of the efficacy of Cayley's parametrization
in lattice perturbation theory is to compute the same background
field one-loop effective action taking into account a resummation of tadpoles.

\def\mn#1{{#1}_{\mu\nu}}
\def\cE{{\cal E}}
In the background field approach, we parametrize the link variables as
$$U_\mu(x) \equiv C(\alpha_\mu)U^0_\mu(x),$$
where $C(x) \equiv (1+iax/2)/(1-iax/2),$ $\alpha$ are the fluctuations in
the gauge potential, and $U^0$ is the background configuration of the
link variables, supposed to be weak ({\it i.e.} close to the identity) and
varying only over long distances.  The lattice covariant derivatives $D^0_\mu$
are now defined with parallel transport by $U^0.$
It is a standard exercise to show that
$$\tr W_{\mu\nu} = \tr C(\mn X)W^0_{\mu\nu}, {\rm with} \qquad
W^0_{\mu\nu}\equiv  U^0_\mu(x)U^0_\nu(x+e_\mu)
U^0_\mu(x+e_\nu)U^0_\nu(x),$$
and
$${\mn{X}\over a} \equiv D^0_\mu\alpha_\nu -D^0_\nu\alpha_\mu + {i\over
2}\left(
2[\alpha_\mu,\alpha_\nu] - a[D^0_\nu\alpha_\mu,\alpha_\mu]
+a [D^0_\mu\alpha_\nu,\alpha_\nu]- a^2[D^0_\nu\alpha_\mu,
D^0_\mu\alpha_\nu]\right),$$
which we divide into two parts, $\mn X/a \equiv 2(\mn E+i\mn B).$
We write $2W^0_{\mu\nu}\equiv G_{\mu\nu} + i H_{\mu\nu},$ with $G=2+O(F^2),$
$H=O(F),$ and
note that $X^2 = 4a^2 (E^2 -B^2 +i\{E,B\}).$
Further,
$$\eqalign{4\Re\tr W_{\mu\nu}= \tr\Bigg[{1\over 1+a^2{\mn X^2}/4}
&\Big(\left(1-i
{{{\mn X}a}\over2}\right)^2(\mn G-i\mn H)\cr
+&\left(1+i{{{\mn X}a}\over2}\right)^2(\mn G+i\mn H)\Big)\Bigg],\cr}$$
so for a one-loop computation we need only
$$2\Re\tr W_{\mu\nu}= \tr\left[{1\over 1+a^2{\mn E^2}}
\left((1-a^2{\mn E^2})\mn G -4ia\mn B\mn H
\right)\right].$$
Note that at this order in the weak-coupling
background field calculation we do not
need to consider the lattice functional Haar measure.

Expanding $S_W$ about $U^0,$
we have
$$S_W = {1\over g^2} \sum_{\rm plaquettes} \Re
\tr\left( 1-W^0_{\mu\nu}\right)
+\Re\tr\left( W^0_{\mu\nu}-W_{\mu\nu}\right),$$
which gives
$$S_W = {1\over g^2} \sum_{\rm plaquettes} \Re
\tr\left( 1-W^0_{\mu\nu}\right)
+\tr\left[{a^2{\mn E^2}\over 1+a^2{\mn E^2}} G
+ {i}{{a\mn B\mn H}\over 1+a^2{\mn E^2}}\right].$$
Define
$\mn\cE=(a^2{\mn E^2}-C)/(1+C).$
with $C$ a number that we shall fix momentarily.
In terms of $\mn\cE,$
$$\eqalign{S_W = &{1\over g^2}\big({1-C\over1+C}\big)
\sum_{\rm plaquettes} \Re
\tr\left(1- W^0_{\mu\nu}\right) \cr
&+{1\over {g^2(1+C)}}\sum_{\rm plaquettes}
\tr\Big[{2a^2{\mn E^2}\over {1+{\mn \cE}}}
+{{\mn\cE}\over1+\mn\cE}(\mn G-2) -i{{a\mn B\mn H}\over1+\mn\cE}\Big].\cr}$$
Observe that the coupling constant $g^2$ in front of the classical
action has been renormalized to $g^2(1+C)/(1-C).$

We now wish to fix $C.$  To this end, rescale the fluctuation fields
by a factor of $g\sqrt{1+C}.$
The part of $S_W$ governing
fluctuations becomes
$$S(\alpha) =  \sum_{\rm plaquettes}\tr\Big[{2a^2{\mn E^2}\over
{1+g^2{\mn \hcE}}}
+{{ \mn\hcE}\over1+g^2\mn\hcE}{1\over1+C}(\mn G-2) -ia{{\mn B\mn H}\over1+
g^2\mn\hcE}\Big],$$
with
$$\mn\hcE= a^2{\mn E^2}-{C\over g^2(1+C)}.$$
It follows therefore that at this order, we should choose
$$C ={ {g^2\langle a^2\mn E^2\rangle}\over {1-g^2\langle a^2\mn E^2\rangle}}
,$$
which implies that
$${1-C\over1+C} = 1-2g^2\langle a^2\mn E^2\rangle.$$
The gauge fixing terms are as usual, with the gauge parameter dependent
on $C$ for Feynman gauge, but this is not a problem.

It should be noted that there are two types of contributions to the
one-loop renormalization of the lattice coupling constant, those that
have continuum analogues and those that are specific to lattice gauge
theory.  Ideally, one would want to incorporate all the lattice specific
renormalizations into a resummed action---the resummation proposed above
does {\it not} include all the lattice specific renormalizations, it only
includes the tadpole diagram that provides the bulk of the renormalization.
Within the present calculation, it does not seem possible to resum the
other type of lattice specific
renormalizations, essentially because they are not
tadpole diagrams.  It may be possible to eliminate such contributions
if one starts with an action different from the Wilson action.

When $\beta=6,$ for SU(3), this tells us that the
coupling constant in front of the classical action is
$$\tilde g^2 = {4\over 3} g^2,$$
at this order in perturbation theory.
This should be compared with the non-perturbative value found numerically
by Lepage and Mackenzie,
$$\tilde g^2 \approx 1.7 g^2.$$
It would appear then that the use of the Cayley parametrization combined with
our resummation of the perturbation expansion is a decent approximation to
the numerical renormalization, since this value of $\beta$ corresponds to
$g^2=1.$ Note that we
have ignored terms of order $g^2\alpha^2$ from the measure[\has]
for example, and we have not included any Symanzik-type improvements
in the action---nevertheless, we still obtained a lowest order result that
is about half of the numerical non-perturbative value.

Two more comments are in order.  Firstly, perturbation theory is still
organized in powers of $g^2$ since higher terms in the fluctuation field
appear in the combination $g^2\hcE.$  Of course, if one calculated at
higher order, one would also need to change the definition of $C$ to be
consistent.  Secondly, the tadpole contribution to $\Lam_L/\Lam_{PV}$
does not appear since $\mn G - 2 $ couples to $\hcE,$ and
$\langle \hcE\rangle=0.$  The one-loop renormalization of the tree action
is still non-trivial since we have only resummed some of the graphs
contributing to the determinant.  In particular, $C$ should be
adjusted order by order in the perturbative expansion in order to eliminate
the largest contribution from diagrams specific to lattice
gauge theory.

\bigskip
\centerline{4. Conclusions}
\bigskip
It is
plausible, on the basis of the simple calculation presented here,
that combining the Cayley parametrization with Symanzik-type
improvements will lead to a systematic lattice perturbation theory that is
much closer to the continuum and to the non-perturbative numerical
simulations.
It seems unlikely that lattice perturbation theory, even if
it is reformulated as described above,  will become a practical
tool for calculations in gauge theories.  The interest in understanding
the structure of lattice perturbation theory is in being able to
match numerical results with continuum results.  As I hope is evident
from the preceding discussion, there is freedom in the definition
of the lattice perturbation expansion that can be used to
render such a comparison more transparent.  Deviations from perturbation
theory can then be understood as either lattice artifacts, or genuine
non-perturbative effects.  The numerical phenomenology of genuine
non-perturbative physics would be useful in constructing analytical
approximations to gauge theory dynamics.  While such systematic
study does not have the theoretical allure of direct attempts
at phenomenological models of low-energy dynamics in gauge theories,
keep in mind the words of Simone Weil---`We must prefer real
hell to an imaginary paradise.'
\bigskip
I am grateful to Mark Alford for comments on the manuscript.
\bigskip
\centerline{References}
\bigskip
\item{\wil.} G.P. Lepage and P.B. Mackenzie, {\sl Phys. Rev.} {\bf D48} (1993)
2250

\item{\wz.} K. Symanzik, {\sl Nucl. Phys.} {\bf B226} (1983) 187

\item{\gross.} R. Dashen and D.J. Gross,
{\sl Phys. Rev.} {\bf D23} (1981) 2340

\item{\z.} M. Alford, W. Dimm, G.P. Lepage, G. Hockney, and P.B. Mackenzie,
{\it QCD on coarse lattices}, in the
Proceedings of {\it Lattice '94}, Bielefeld, Germany.

\item{\spglcs.} G.P. Lepage, seminar at Princeton University, April 1995

\item{\has.} A. Hasenfratz and P. Hasenfratz, {\sl Phys. Lett.} {\bf 93B}
(1980) 165

\end